# Dynamical Complexity, Intermittent Turbulence, Coarse-Grained Dissipation, Criticality and Multifractal Processes

Tom Chang* and Cheng-chin Wu**

*MIT Kavli Institute for Astrophysics and Space Research, Cambridge, MA 02139 USA
**UCLA Institute of Geophysics and Planetary Physics, Los Angeles, CA 90095 USA

**Abstract.** The ideas of dynamical complexity induced intermittent turbulence by sporadic localized interactions of coherent structures are discussed. In particular, we address the phenomenon of magnetic reconfiguration due to coarse-grained dissipation as well as the interwoven connection between criticality and multifractal processes. Specific examples are provided.

## DEFINITION OF DYNAMICAL COMPLEXITY

We start this short discourse with the definition of dynamical complexity. "Dynamical complexity" is a phenomenon exhibited by a nonlinearly interacting dynamical system within which multitudes of different sizes of large scale "coherent structures" are formed, resulting in a global nonlinear stochastic behavior for the dynamical system, which is vastly different from that could be surmised from the original dynamical equations [1].

## COARSE-GRAINED DISSIPATION AND MAGNETIC RECONFIGURATION

We now proceed to demonstrate that the phenomenon of intermittent turbulence in magnetized space plasmas is a phenomenon of dynamical complexity. Plasmas are known for their ability to form multitudes of coherent structures of varied sizes with scales that are generally much larger than those of the constituent particles (ions, electrons, neutrals) of the plasma medium. Examples of plasma coherent structures are flux tubes, drift vortices, Alfvén vortices, convective structures, stream tubes, double layers and various types of solitons such as electron and ion holes. In the magnetotail, solar wind and magnetopause, for example, one dominant form of the coherent structures is the Alfvénic flux tubes [2-5].

When coherent Alfvénic flux tubes with the same polarity migrate toward each other, strong local magnetic shears are created, Fig. 1. It has been demonstrated by Wu and Chang [6] and Chang et al. [5] that the existing sporadic nonpropagating fluctuations in the strong local shear region, particularly those close to the neutral sheet (i.e., at the location where the local magnetic field vanishes), generally will stay in the region and continue to interact nonlinearly. Fluctuations away from the neutral sheet region, on the other hand, are nonresonant and will therefore propagate away along magnetic field lines as Alfvén waves. Combined with the magnetic shear geometry, the resonant fluctuations will induce a nonlinear instability near the neutral sheet region, which will produce more fluctuations; some of these are nonresonant and the remaining resonant. The nonresonant fluctuations will again propagate away as Alfvén waves while the resonant ones will join the other resonant fluctuations and interact nonlinearly; thereby, broadening the resonance region. This combined phenomenon of "coarse-grained dissipation" depletes the energy originally contained in the coarse-grained magnetic fields near the shear region, and eventually initiates a reconfiguration of the topologies of the coherent structures of the same polarity by breaking some of the closed field lines of each of the coherent structures and then reconnects them into single closed field lines. And, this process continues with the system adjusting continually with the surrounding environment until all free energies are exhausted; eventually, leading to the formation of a single combined coherent structure with one set of concentric closed field lines, Fig. 1. The final state of the resulting coherent structure will have less average energy due to the combined dissipation of Alfvénic propagation of nonresonant fluctuations and nonlinear interactions of the resonant fluctuations (i.e., resonance broadening).

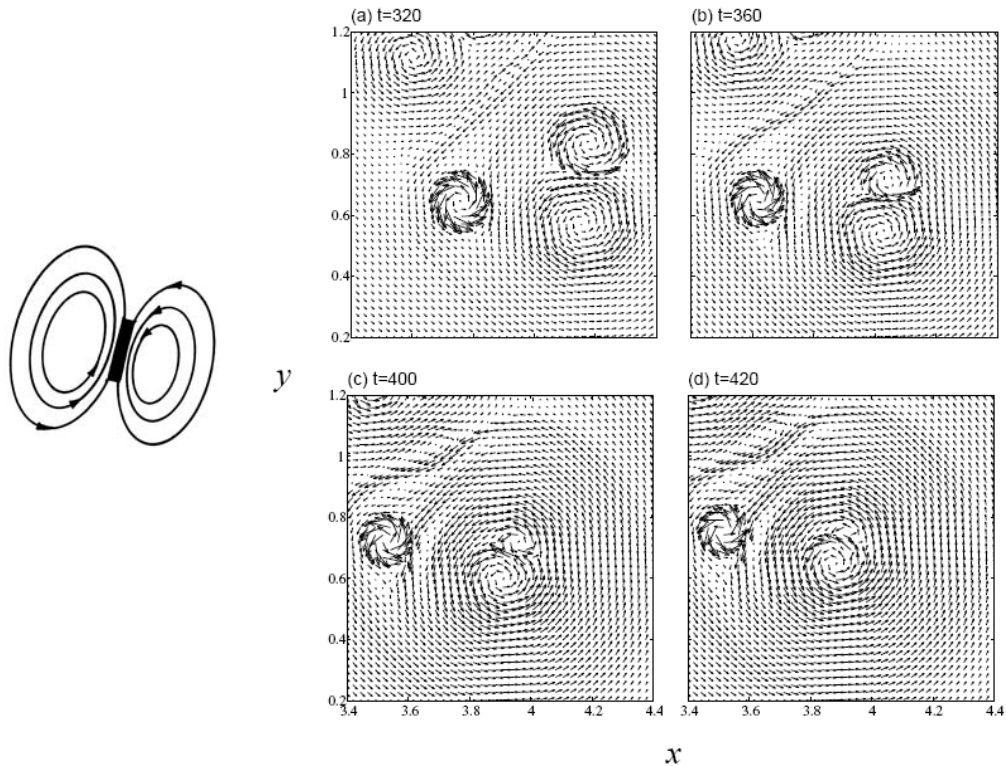

**FIGURE 1.** Cross-sectional views of Alfvénic flux tubes. Left: Schematic of merging. Blackened area indicates location of strong shear. Right: 2D MHD simulation of merging. Arrows indicate magnitude and direction of magnetic fields. Adapted from [5].

This merging or "magnetic reconfiguration" process may repeat over and over again among the coherent structures of the same polarity with all scales. On the other hand, when coherent structures of opposite polarities approach each other due to the forcing of the surrounding plasma, they might repel each other, scatter, or induce magnetically quiescent localized regions.

Topological reconfigurations of such nature occur quite frequently for the dynamical interactions of coherent structures in space plasmas and are not limited just to flux tubes; e.g., Sundkvist et al. [7] and Alexandrova et al. [8] have observed intermittent interactions of drift Alfvén vortices in the cusp and magnetosheath, respectively. Similarly, such enhanced intermittency at the intersection regions of whistler coherent structures has also been surmised by Consolini et al. [9] in the plasma sheet. All such intermittent interactions are quite akin to the avalanche phenomenon prevalent in sandpile models [10, 11]. They are the origin of the various observed magnetic reconnection signatures in space plasmas.

This stochastic behavior of the interactions of the plasma coherent structures is a phenomenon of "dynamical complexity" as defined above.

## HIGH PROBABILITY OF LARGE EVENTS

Because the coherent structures are numerous and outsized, we expect the fluctuations within the interaction regions of these structures (resonance overlap regions) are generally large and can occur relatively often than those that would have been expected from a medium of uniformly sized plasma particles. Such statistical characteristic of high probability of large fluctuations at small scales exemplifies the phenomenon of intermittent turbulence in space plasmas.

## MULTIFRACTALS

As we have seen above, turbulence in space plasmas generally encompass fluctuations of all varieties and sizes, which interact and propagate throughout the plasma medium. For illustrative purposes, let us visualize some particular fluctuations that have conventional geometrical properties in a three-dimensional Euclidean space. Because of their sporadic and localized nature, it is easy to imagine that they generally cannot fill the full three-dimensional space that they occupy at a given time. Or, the dimension these fluctuations occupy is less than 3. Such geometrical property was popularized by Mandelbrot [12] when he coined the word, "fractals".

Actual fluctuations in plasma turbulence generally do not have the conventional geometrical properties. We must then devise some abstract "measure" that characterize the properties of the fluctuations and evaluate their fractal characteristics that may be interpreted with geometrical analogs. Consider, for example, the spatial series of the fluctuations of the strength of the magnetic field, $B(x_i)$, from a 2D MHD simulation of fully developed intermittent turbulence along some constant value of $y$ at time $t$, where $x_i = i\delta$ with $i = 0, 1, 2, ...., N$ and $\delta$ the length between grid points. From this series, we can construct a coarse-grained moment measure of the fluctuations, "the *qth order* structure function", as follows

$$S_q(\delta B^2, \Delta) = \left\langle \left| B^2(x_i + \Delta) - B^2(x_i) \right|^q \right\rangle \tag{1}$$

where $<\cdots>$ represents the ensemble average. The motivation here is that for intermittent turbulence different moments emphasize different peaks in the fluctuating series. For small $\Delta$, $S_q$ may vary with the coarse-graining as a power law: $S_q \approx \Delta^{\zeta_q}$. If it does, the exponent $\zeta_q$ might be considered as a dimension of $S_q$. Generally $\zeta_q$ is an irrational number and cannot be surmised from dimensional analysis. Thus, it may be visualized as a fractal dimension of the *qth* order structure function. If $\zeta_q = \zeta_1 q$, then the fractal property of the fluctuating series is fully characterized by the value of $\zeta_1$. Such fluctuations are then said to be "monofractal" or "self-affine", i.e., the fractal characteristics for all the moment orders are similar to each other. For general intermittent turbulence, on the other hand, $\zeta_q$ would be a nonlinear function of $q$, or "multifractal". Calculations similar to such have been carried out for the solar wind [e.g., 13-18] and elsewhere in the space environment [e.g., 19-25]. Figure 2 displays the multifractal characteristics of a 2D MHD simulation.

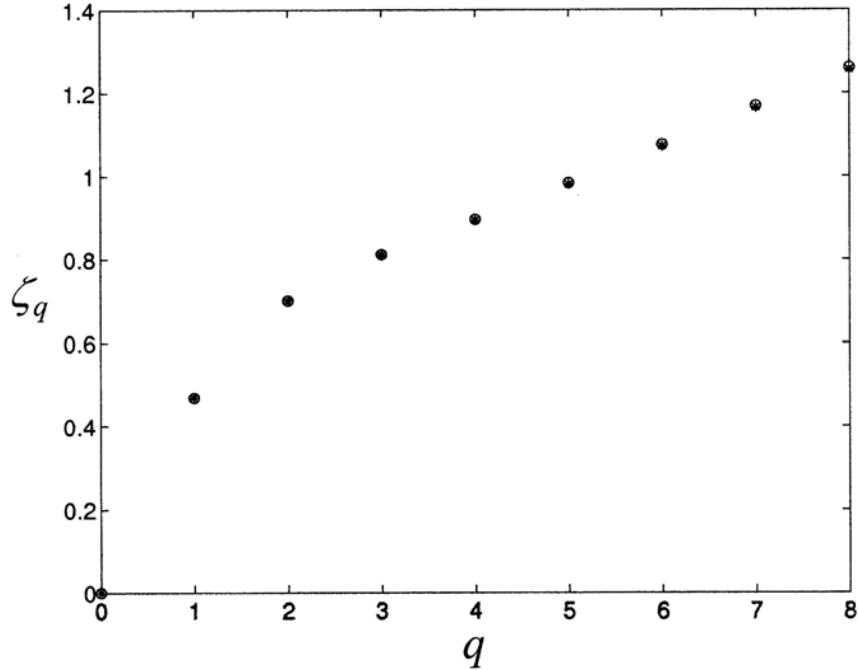

**FIGURE 2.** Multifractal structure function exponents from a 2D MHD simulation.

## FROM CRITICALITY TO INTERMITTENCY

For nonlinear stochastic systems exhibiting complexity, the correlations among the fluctuations of the random dynamical fields (electric, magnetic and velocity fields, etc.) are generally very long-ranged and there exist many correlation scales (as exhibited by the various outsized plasma coherent structures). The dynamics of such systems (dynamical criticality) are notoriously difficult to handle either analytically or numerically. On the other hand, since the correlations are extremely long-ranged, it is reasonable to expect that the system will exhibit some sort of invariance property under coarse-grained scaling transformations [26]. The mathematical consequence of this invariant theory (the dynamic renormalization group) is that, close to dynamical criticality, certain linear combinations of the parameters that characterize the stochastic phenomenon will correlate with each other in the form of power laws.

Power law behavior has been detected in the probability distributions of solar flares [27, 28], in the AE burst occurrences as a function of the AE burst strength [29], in the global auroral UVI imagery of the statistics of size and energy dissipated by the magnetospheric system [30], in the probability distributions of spatiotemporal magnetospheric disturbances as seen in the UVI images of the nighttime ionosphere [31], and in the probability distributions of durations of Bursty Bulk Flows [32]. Such invariant behavior has been called self-organized criticality (SOC [10]), or more preferably forced and/or self-organized criticality (FSOC, [26]).

We now ask a deeper question: How does the invariant behavior of the structure functions that changes with moment order for intermittent turbulence amalgamate with the concept of criticality (SOC or FSOC) outlined above?

For a stochastic process that is monofractal, the structure function exponents vary with the moment order linearly. Thus a single invariant, such as $I_1 = S_1 / \Delta^{\zeta_1}$ would suffice to characterize the turbulent behavior for fully developed turbulence that is self-affine. On the other hand, for multifractal stochastic processes, the full range of scale invariants, such as $I_q = S_q / \Delta^{\zeta_q}$ must be addressed in order to understand the stochastic behavior near criticality, which is generally the case for intermittent turbulence.

Probability distribution functions (PDF) of most of the observed and simulated results for intermittent turbulence generally have shapes very close to that of a Gaussian at small fluctuations. Since observational or numerical results obtained for SOC/FSOC generally result from statistical evaluations related to low order moments which are characterized mainly by those portions of the PDFs for small fluctuations, the obtained invariant scaling properties for multifractal intermittent turbulence should be quite close to those that would be expected from Gaussian distributions, given the uncertainties of the observations and numerical simulations. And, such observed scaling exponents would satisfy the standard scaling relations of SOC/FSOC. This is recently demonstrated for solar flare statistics [28] and also numerically [33].

The PDFs for intermittent turbulence will generally deviate more and more from the shape of a Gaussian for larger and larger fluctuations. In addition, as the gaps among the interacting coherent structures become larger, the sporadic localized intense fluctuations will be separated by more and more regions of less activity, leading to the high probability of large fluctuation deviations from the mean. Such "crossover effects" [26] generate the multifractal nature of the higher order structure functions for intermittent turbulence. In fact, we expect to begin to see these effects even for relatively small values of the moment order, $q$. And, this is reflected in the actual observational and numerical simulation results [28, 33]. As a consequence, the corresponding scaling exponents will satisfy more complicated scaling relations.

Thus, the conclusion here is that the conventional concepts of SOC or FSOC as tested by the lower order spatiotemporal statistical averages would generally be satisfied for intermittent turbulence. These ideas, however, will need to be generalized to include "crossover effects" when the concept of invariants, etc. are extended to the high order scaling properties both for the structure functions of higher order and for other higher order correlation and response functions of the spatiotemporal variety.

*To summarize, in this short discourse, we have considered the phenomenon of coarse-grained dissipation and the interwoven concepts of SOC/FSOC and intermittent turbulence from the modern point of view of dynamical complexity.*

## ACKNOWLEDGMENTS

This research is partially supported by AFOSR, NASA, and NSF.